\documentclass[12pt]{article}

\usepackage{amssymb}
\usepackage{amsmath}
\usepackage{latexsym}
\usepackage{epsfig}
\usepackage{graphicx}
\usepackage{subfigure}
\usepackage{wasysym}
\usepackage{threeparttable}
\usepackage{booktabs}
\usepackage[round]{natbib}
\usepackage{color}
\usepackage{epstopdf}
\usepackage{bm}
\usepackage{tabularx}
\usepackage{verbatim}
\usepackage{hyperref}
\usepackage{float}
\usepackage[ruled,vlined]{algorithm2e}
\usepackage{ulem}

\def\boxit#1{\vbox{\hrule\hbox{\vrule\kern6pt
          \vbox{\kern6pt#1\kern6pt}\kern6pt\vrule}\hrule}}

\def\bse{\begin{eqnarray*}}
\def\ese{\end{eqnarray*}}
\def\be{\begin{eqnarray}}
\def\ee{\end{eqnarray}}
\def\bq{\begin{equation}}
\def\eq{\end{equation}}
\def\bse{\begin{eqnarray*}}
\def\ese{\end{eqnarray*}}

\begin{document}
\thispagestyle{empty} \baselineskip=28pt

\begin{center}
{\LARGE{\bf A Bayesian Functional  Data  Model for Surveys Collected under Informative Sampling with Application to Mortality Estimation using NHANES}}
%{\LARGE{\bf A Comprehensive Overview of Unit Level Modeling of Survey Data Under
%    Informative Sampling with Applications to Small Area Estimation}}

\end{center}

\baselineskip=12pt

\vskip 2mm
\begin{center}
Paul A. Parker\footnote{(\baselineskip=10pt to whom correspondence should be
  addressed) Department of Statistics, University of Missouri, 146 Middlebush Hall,
  Columbia, MO 65211-6100, paulparker@mail.missouri.edu}\, and
Scott H. Holan\footnote{\baselineskip=10pt Department of Statistics, University of
  Missouri, 146 Middlebush Hall, Columbia, MO 65211-6100,
  holans@missouri.edu}\,\footnote{\baselineskip=10pt Office of the Associate Director for Research and Methodology, U.S. Census Bureau, 4600 Silver
  Hill Road, Washington, D.C. 20233-9100, scott.holan@census.gov}

\end{center}

\vskip 4mm

\begin{center}
  {\bf Abstract}
\end{center} 
Functional data are often extremely high-dimensional and exhibit strong dependence structures but can often prove valuable for both prediction and inference. The literature on functional data analysis is well developed; however, there has been very little work involving functional data in complex survey settings. Motivated by physical activity monitor data from the National Health and Nutrition Examination Survey (NHANES), we develop a Bayesian model for functional covariates that can properly account for the survey design. Our approach is intended for non-Gaussian data and can be applied in multivariate settings. In addition, we make use of a variety of Bayesian modeling techniques to ensure that the model is fit in a computationally efficient manner. We illustrate the value of our approach through an empirical simulation study as well as an example of mortality estimation using NHANES data.

\baselineskip=12pt 

\baselineskip=12pt
\par\vfill\noindent
{\bf Keywords:}  Functional principal components, Horseshoe prior,  National Health and Nutrition Examination Survey (NHANES), P\'{o}lya-Gamma, Pseudo-likelihood.

\par\medskip\noindent

\clearpage\pagebreak\newpage \pagenumbering{arabic}
\baselineskip=24pt

\section{Introduction}\label{sec: intro}

The use of functional data as either a response or covariate has seen wide usage in recent years. Applications that utilize functional data include longitudinal data analysis \citep{yao2005functional}, ecology \citep{yang2013ecological}, small area estimation \citep{porter2014spatial}, as well as many others. However, typical models for functional data typically assume a sample that is representative of the population, and thus are not directly applicable to many survey datasets, especially under informative sampling. For example, the National Health and Nutrition Examination Survey (NHANES) contains functional data in the form of activity monitor curves, yet the survey design is complex leading to a sample that is not representative of the population.

The literature on functional data analysis for survey data is quite sparse. \citet{savitsky2016bayesian} consider the case of functional responses under informative sampling. They treat the response as a Gaussian process to handle functional dependence while simultaneously using a weighted Bayesian pseudo-likelihood to account for informative sampling. One drawback of this approach is that computation under the Gaussian process formulation can become prohibitively difficult in high-dimensional settings.

More recently, \citet{leroux2019organizing} explore scalar on function regression to predict 5-year mortality rate based on NHANES physical activity covariates. Their approach employs existing software packages that use the survey weights to construct appropriate point estimates but are unable to give correct estimates of uncertainty based on the sample design. The authors state that a resampling procedure may be used to give appropriate standard errors, but the approach is beyond the scope of the paper. Ultimately, their use of scalar on function regression was more exploratory and not intended to fully account for the survey design.

In this work, we develop a Bayesian model for scalar on function regression of survey data under informative sampling. Through the use of a Bayesian pseudo-likelihood \citep{sav16}, we are able to give appropriate measures of uncertainty. In addition, we use data augmentation to ensure that the model can be fit in an efficient manner via Gibbs sampling. We also provide an extension to Multinomial response data, which allows for certain multivariate problems to fit into our framework. Similar to \citet{leroux2019organizing}, we are primarily motivated by the topic of mortality estimation with NHANES physical activity covariates,  though we note that this methodology is generally applicable to any type of functional survey data. The remainder of this work is outlined as follows. In Section \ref{sec: methods} we describe our methodology along with necessary background material. Section \ref{sec: data} outlines the motivating NHANES dataset. In Section \ref{sec: sim} we conduct an empirical simulation study that utlizes the public-use NHANES activity monitor data. We also present a data analysis of the public-use NHANES data in Section \ref{sec: DA}. Finally, we provide concluding remarks and discussion in Section \ref{sec: disc}. All of our relevant code and processed data is available for download at \url{https://github.com/paparker/survey_FDA}.

\section{Methodology}\label{sec: methods}

\subsection{Informative Sampling}
In many survey data settings, there is dependence between a unit's probability of selection and the response of interest. This is termed \textit{informative sampling}  and is known to introduce bias into the model when ignored. Thus, in survey data settings, it is important to account for the survey design in some manner in order to eliminate or reduce this bias. In other words, complex sample designs can lead to samples that are unrepresenative of the population and, thus,  the sample model should be adjusted in some way to account for this.

\citet{par19} give an overview of various methods to account for informative sampling. Of primary interest is the pseudo-likelihood (PL) method introduced by \citet{ski89} and \citet{bin83}. This approach adjusts the likelihood function by exponentially weighting each unit's likelihood contribution by the corresponding survey weight (i.e. the inverse of the selection probability),
\begin{equation}\label{E: PL}
  \prod_{i \in \mathcal{S}}  f( y_i \mid \boldsymbol{\theta})^{w_i},
\end{equation} where $\mathcal{S}$ indicates the sample, $y_i$ represents the response value for unit $i$ with survey weight $w_i$. In a frequentist setting, this PL can be maximized to give a point estimate for $\bm{\theta},$ however more complex procedures are necessary to give appropriate estimates of uncertainty.

\citet{sav16} show that a PL may also be used in a Bayesian framework. In particular, they show that under informative sampling, the use of a PL along with a prior specification leads to a pseduo-posterior distribution,
$$\hat{\pi}(\bm{\theta} | \mathbf{y}, \mathbf{\tilde{w}}) \propto \left\{ \prod_{i \in \mathcal{S}} f(y_{i} | \bm{\theta})^{\tilde{w}_{i}} \right\} \pi (\bm{\theta}),$$
that converges to the population posterior distribution. In this scenario, it is important to scale the weights to sum to the sample size in order to attain the appropriate estimates of uncertainty. These scaled weights are represented by $\tilde{w}_i.$ This formulation applies generally to Bayesian models and is the approach we use herein to account for informative sampling.

\subsection{Non-Gaussian Data}\label{sec: ng data}

Modeling non-Gaussian data types in a Bayesian setting can be computationally burdensome, especially while accounting for informative sampling. \citet{parker2020computationally} utilize a data augmentation approach to construct a flexible mixed model for Binomial and Multinomial data under informative sampling. Their model for Binomial data is given by,
    \begin{equation}\label{eq: mod1}
    \begin{split}
        \bm{Z} | \bm{\beta, \eta} & \propto \prod_{i \in S} \hbox{Bin}\left(Z_i | n_i, p_i \right)^{\stackrel{\sim}{w}_i} \\
        \hbox{logit}(p_i) &= \bm{x}_i' \bm{\beta} + \bm{\phi}_i' \bm{\eta} \\
        \bm{\eta}|\sigma^2_{\eta} & \sim \hbox{N}_r(\bm{0_r}, \sigma_{\eta}^2 \bm{I}_r ) \\
        \bm{\beta} & \sim  \hbox{N}_q(\bm{0_q}, \sigma_{\beta}^2 \bm{I}_q ) \\
        \sigma_{\eta}^2 & \sim \hbox{IG}(a, b) \\
        & \sigma_{\beta},  a, b >0,
    \end{split} 
\end{equation} where $Z_i$ represents the response value for unit $i$ in the sample. In this case, $\bm{x}_i$ is a vector of fixed effects covariates and $\bm{\phi}_i$ represents a set of spatial basis functions.

In order to fit this model in a computationally efficient manner, Pólya-Gamma data augmentation is used. Specifically, letting $\hbox{PG}(\cdot,\cdot)$ represent a Pólya-Gamma distribution, \citet{pol13} show that 
\begin{equation*}\label{eq: pg}
   \frac{(e^{\psi})^a}{(1 + e^{\psi})^b} = 2^{-b}e^{\kappa \psi} \int_0^{\infty} e^{-\omega \psi^2/2} p(\omega) d\omega, 
\end{equation*} where $\kappa = a - b/2$ and $p(\omega)$ is a $\hbox{PG}(b,0)$ density. They further show that $(\omega | \psi) \sim \hbox{PG}(b,\psi)$. The PL in (\ref{eq: mod1}) can be written,
\begin{equation*}\label{eq: PL_bin}
   \prod_{i \in \mathcal{S}} \left(\frac{(e^{\psi_i})^{Z_i}}{(1 + e^{\psi_i})^{n_i}}\right)^{\tilde{w}_i} =
           \prod_{i \in \mathcal{S}} \frac{(e^{\psi_i})^{Z_i^*}}{(1 + e^{\psi_i})^{n_i^*}},
\end{equation*} where $\psi_i = \hbox{logit}(p_i),$ $Z_i^*=Z_i \times \tilde{w}_i,$ and $n_i^*=n_i \times \tilde{w}_i$. This allows for data augmentation of a latent Pólya-Gamma random variable that leads to conjugate Normal priors on the regression parameters.

The Binomial model in (\ref{eq: mod1}) can also be extended to Multinomial or Categorical data. Following \citet{linderman15}, the Multinomial distribution with $C$ categories can be rewritten as
\begin{equation*}
    \begin{split}
        \hbox{Multinomial}(\bm{Z}|n, \bm{p})=\prod_{c=1}^{C-1} \hbox{Bin}(Z_c|n_c, \tilde{p}_c),
    \end{split}
\end{equation*} where
\begin{equation*}
    n_c = n - \sum_{j < c}Z_j, \; \; \tilde{p}_c = \frac{p_c}{1 - \sum_{j<c} p_j}, \; \; c=2,\ldots,C.
\end{equation*} In this light, a series of $C-1$ Binomial models may be fit to estimate the parameters for a Multinomial data model.

The modeling framework of \citet{parker2020computationally} is useful for fitting Binomial data under informative sampling, such as the NHANES mortality data of interest; however, the approach must be extended in order to consider functional covariates.

\subsection{Functional Covariates}

Consider the case where we have $J$ functional covariates and $\kappa_{ij}(t), t \in \mathcal{T}$ denotes the $j$th functional covariate ($j=1,\ldots,J$) for unit $i$ at time $t.$ In our case, the domain is time, though other domains may be appropriate depending on the type of functional data. Then, (\ref{eq: mod1}) can be extended for functional covariates by letting
$$
\hbox{logit}(p_i) = \bm{x}_i'\bm{\beta} + \sum_{j=1}^J \int_{\mathcal{T}}\eta_j(t)\kappa_{ij}(t) dt,
$$ where $\eta_j(t)$ is a functional regression parameter associated with functional covariate $j.$ In what follows, we will assume $J=1$ (and drop the subscript $j$), as is the case in our example, but we note that the approach is still applicable for $J>1.$

In order to reduce the dimension of the problem, we can use a basis expansion representation. In particular, let $\left\{\phi_k(t):k=1,2,\ldots\right\}$ be a complete orthonormal basis of the domain $\mathcal{T}.$ Then, we can represent the functional covariate as
$$
\kappa_i(t)=\sum_{k=1}^{\infty}\xi_i(k)\phi_k(t)
$$ and 
$$
\eta(t)=\sum_{k=1}^{\infty}b(k)\phi_k(t),
$$ where $\xi_i(k)$ and $b(k)$ are the expansion coefficients for $\kappa_i(\cdot)$ and $\eta(\cdot)$ respectively. Now, appealing to orthonormality,
\begin{equation}\label{eq: func}
    \hbox{logit}(p_i) = \bm{x}_i'\bm{\beta} +  \int_{\mathcal{T}}\eta(t)\kappa_{i}(t) dt = \bm{x}_i'\bm{\beta} + \sum_{k=1}^{\infty}b(k)\xi_i(k).
\end{equation} Note that any orthonormal basis may be used here, though we use functional principal components selected through the fast covariance estimation (FAST) approach \citep{xiao2016fast}. This is easily implemented via the use of the \texttt{refund} package in \texttt{R} \citep{refund}.

In practice, the summation in (\ref{eq: func}) is truncated to $K$. For our purposes, we truncate the summation (i.e., choose K) such that the retained components explain 95\% of the variation in the functional data. This results in a finite, though potentially large, number of basis functions. Furthermore, any given basis function may not necessarily be related to the response. Thus, we require some form of variable selection and shrinkage estimator. By doing so, the variable selection prior is able to determine which components of the variation in functional data are correlated with the response. This is similar to the approach taken by \citet{holan2010modeling}.

\subsection{Horseshoe Prior}

In order to provide shrinkage to our functional regression coefficients, we utilize the Horsehoe prior introduced by \citet{carvalho2010horseshoe}. Although many other methods of Bayesian variable selection exist, this has the advantage of being fully specified, without requiring hyperparameter selection, as well as providing minimial shrinkage to strong signals while still providing a high degree of shrinkage for noise. To implement the Horsehoe prior for (\ref{eq: func}), we use the following hierarchy,
\begin{align*}
    b(k)|\lambda_k,\tau & \stackrel{ind}{\sim} \hbox{N}(0,\lambda_k^2 \tau^2), \; k=1,\ldots,K \\
    \lambda_k & \stackrel{ind}{\sim} \hbox{C}^+(0,1) \\
    \tau & \sim \hbox{C}^+(0,1),
\end{align*} where $\hbox{C}^+(\cdot,\cdot)$ represents the Cauchy density truncated below at zero. This prior is considered a global-local shrinkage approach. This can be seen by recognizing that $\tau$ applies to all regression parameters and determines the overall level of shrinkage, whereas $\lambda_k$ is local and applies to a specific coefficient. In this way, coefficients corresponding to noise can attain a higher degree of shrinkage than those with strong signals.

The half-Cauchy priors used in the Horsehoe are not conjugate. However, \citet{makalic2015simple} use a data augmentation approach to allow for Gibbs sampling within the Horshoe prior framework. In particular,they use a scale mixture representation of the half-Cauchy such that when $x \sim \hbox{C}^+(0,A)$, then $x^2 | a \sim \hbox{IG}(1/2, 1/a)$ and $a \sim \hbox{IG}(1/2, 1/A^2),$ where $\hbox{IG}(a,b)$ represents the Inverse Gamma distribution with shape parameter $a$ and scale parameter $b.$ This leads to an alternate formulation of the Horsehoe prior hierarchy,
\begin{align*}
    b(k)|\lambda_k,\tau & \stackrel{ind}{\sim} \hbox{N}(0,\lambda_k^2 \tau^2), \; k=1,\ldots,K \\
    \lambda_k^2| \nu_k & \stackrel{ind}{\sim} \hbox{IG}(1/2,1/\nu_k) \\
    \tau^2| \nu_{\tau} & \sim \hbox{IG}(1/2,1/\nu_{\tau}) \\
    \nu_1,\ldots,\nu_K, \nu_{\tau} & \stackrel{ind}{\sim} \hbox{IG}(1/2,1),
\end{align*} that allows for straightforward Gibbs sampling.

\subsection{Functional Data Model under Informative Sampling}

We now present our model for non-Gaussian data under informative sampling with functional covariates, which makes use of the modeling elements discussed so far:
    \begin{equation}\label{eq: mod2}
    \begin{split}
        \bm{Z} | \bm{\beta, \eta} & \propto \prod_{i \in S} \hbox{Bin}\left(Z_i | n_i, p_i \right)^{\stackrel{\sim}{w}_i} \\
        \hbox{logit}(p_i) &= \bm{x}_i' \bm{\beta} + \sum_{k=1}^{K}b(k)\xi_i(k) \\
        \bm{\beta} & \sim  \hbox{N}_q(\bm{0_q}, \sigma_{\beta}^2 \bm{I}_q ) \\
            b(k)|\lambda_k,\tau & \stackrel{ind}{\sim} \hbox{N}(0,\lambda_k^2 \tau^2), \; k=1,\ldots,K \\
    \lambda_k^2| \nu_k & \stackrel{ind}{\sim} \hbox{IG}(1/2,1/\nu_k) \\
    \tau^2| \nu_{\tau} & \sim \hbox{IG}(1/2,1/\nu_{\tau}) \\
    \nu_1,\ldots,\nu_K, \nu_{\tau} & \stackrel{ind}{\sim} \hbox{IG}(1/2,1) \\
        & \sigma_{\beta}^2 >0.
    \end{split} 
\end{equation} In this model, $\bm{x}_i$ represents a $q$-dimensional vector of scalar covariates and $\xi_i(k)$ represents the $k$th basis expansion coefficient for observation $i.$ 

This model makes use of a Bayesian pseudo-likelihood to account for informative sampling, allowing for population level inference. We also make use of Pólya-Gamma data augmentation for efficient Gibbs sampling. If desired, prior information on the scalar covariates may be incorporated through the selection of $\sigma^2_{\beta},$ though we use a relatively diffuse prior by letting  $\sigma^2_{\beta}=10.$ The full conditional distributions are given in the Appendix.

It is straightforward to implement the model in the Multinomial data setting through the use of a stick-breaking representation, as discussed in Section \ref{sec: ng data}. It would also be straightforward to use a Gaussian pseudo-likelihood in place of the Binomial one given here, as conjugacy would be retained.

\section{NHANES Data Description}\label{sec: data}
The NHANES is a survey conducted by the National Center for Health Statistics that utilizes a complex survey design to collect health and nutrition data in the United States. Of primary interest to us is the physical activity monitor (PAM) data collected during the 2003-2004 and 2005-2006 samples. Along with this, we are interested in mortality as a response value.

NHANES provides microdata to the public, however, a substantial amount of data processing is required to utilize the data for inference. \citet{leroux2019organizing} provide very helpful exposition on processing the data as well as the \texttt{rnhanesdata} package in \texttt{R} for doing so. All analyses in this work were conducted using data that was prepared and processed in the same manner as \citet{leroux2019organizing}.

In particular, we use the NHANES samples from 2003-2004 and 2005-2006, as these contain the PAM data of interest. For each minute during a seven consecutive day period, the data contains an activity intensity value for each subject. Not all subjects were in compliance, resulting in variation in wear-time between subjects. Thus, all subjects that had less than 3 days of 10 hours or greater wear-time were dropped. In addition, subjects outside of the age range 50-85, or who were missing mortality or age data were also dropped. The resulting sample size was 3,208. In addition to PAM data as a functional covariate, we use age as a scalar covariate.

The PAM activity measurements were transformed using $f(x)=\hbox{log}(1+x).$ The subjects in the sample had a varying numbers of days with activity data. To account for this, we use the PAM data averaged across days within subjects, resulting in a single 24 hour curve for each subject.

The NHANES sample contains the required survey weights. Following \citet{leroux2019organizing}, these are reweighted to account for missing data. More in depth reweighting schemes may be desired in practice. However, discussion on reweighting procedures for missing data is beyond the scope of this work. Thus, for illustration, we do not consider this problem further.

\section{Empirical Simulation Study}\label{sec: sim}
The goal of our simulation is twofold. First, we want to confirm that the model is able to adjust for an informative sampling mechanism in order to allow for population level inference. Second, we want to assess whether or not the use of functional covariates, as given in the model, leads to improved estimates for units in the population.

To design such a simulation, we begin by treating the existing NHANES sample data as our population. This provides a baseline truth for which we can compare to. Next, we subsample from the NHANES data in an informative manner. Doing so, we are able to fit the model using the subsampled data and then compare to the population truth (i.e. the original sample data). To take this subsample, we use probability proportional to size sampling via the Poisson method \citep{brewer1984poisson} with an expected sample size of 500. We construct the size variable as $s_i=\hbox{exp}\left\{w_i^* + 2*I(Z_i=1)\right\}$ where $w_i^*$ is the NHANES reported survey weight after scaling to have mean zero and variance 1, and $I(Z_i=1)$ is an indicator that the $i$th respondent died within 5 years of the survey. Through this subsampling procedure, we obtain a new set of weights that are the inverse probabilities of selection. 

The response data of interest is the binary indicator of 5-year mortality. We use age along with an intercept term as a scalar covariate and the PAM data as a functional covariate. After subsampling, we fit our functional data model (FM-W) from  (\ref{eq: mod2}). We also fit a basic version of the model that uses only the scalar covariate and disregards the functional data (SM-W). In addition we implement unweighted versions of each of these models (FM-UW and SM-UW).

After fitting each of these models, we are able to make mortality predictions for the entire population. This results in a 5-year probability of mortality for each person in the population that we can compare to their actual mortality result. Because these outcomes are binary, we use binary cross-entropy (BCE) as a loss function to compare rather than mean-squared error. This is calculated as,
$$
\hbox{BCE}=-\frac{1}{N}\sum_{i=1}^N Z_i \hbox{log}(\hat{p}_i) + (1-Z_i)\hbox{log}(1-\hat{p}_i),
$$ where $\hat{p}_i$ is the posterior mean probability of mortality for unit $i=1,\ldots,N$ in the population with size $N.$ Note that a lower
value of BCE indicates a better model fit.

\begin{figure}[H]
    \begin{center}
        \includegraphics[width=150mm]{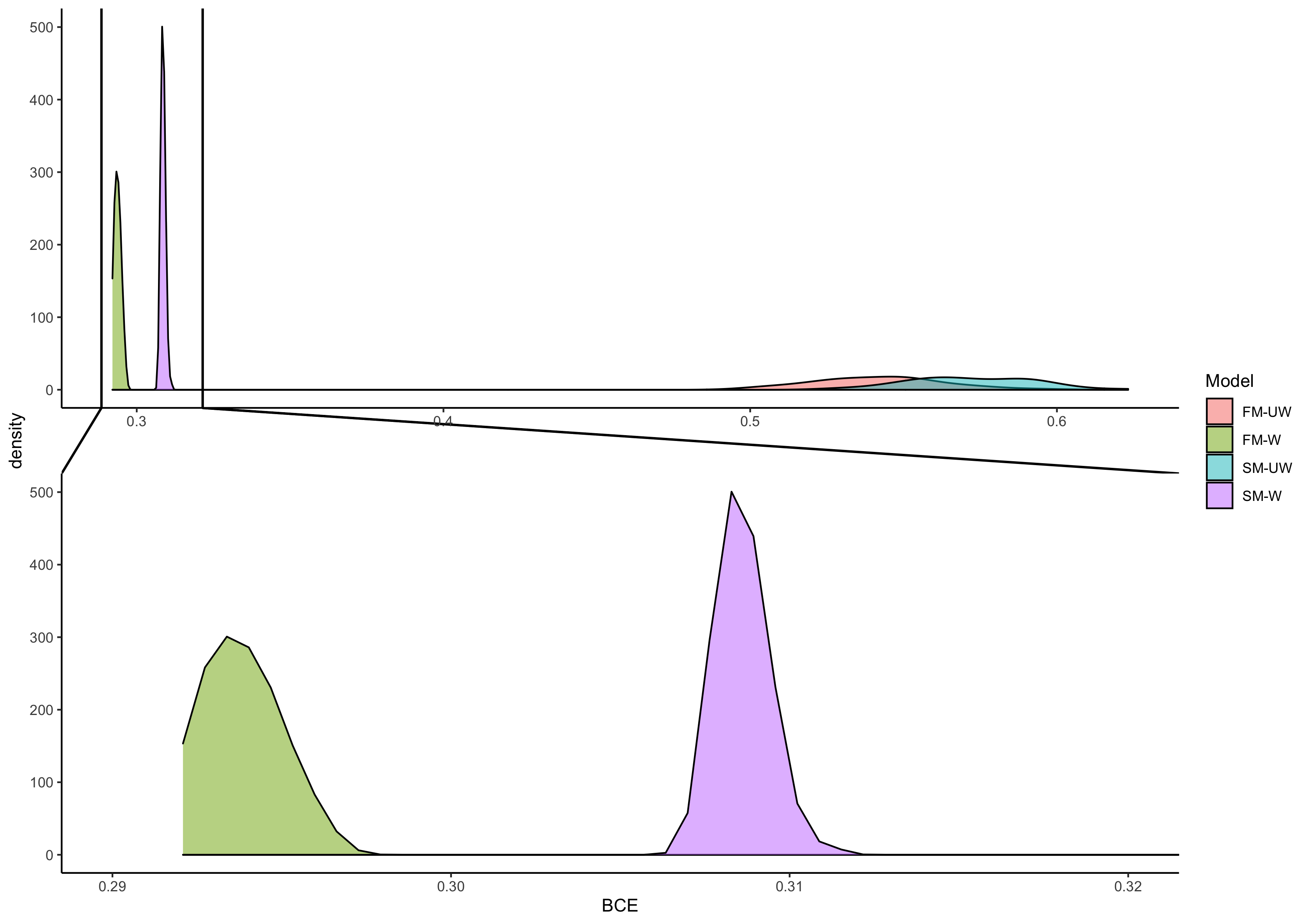}
        \caption{\baselineskip=10pt Simulation based distribution of BCE values under each model. The lower subplot focuses on the two weighted models which have substantially lower BCE.}
        \label{fig:bce}
    \end{center}               
\end{figure}

We repeat this subsampling and model fitting procedure 50 times, resulting in a distribution of loss BCE values under each model. We compare these distributions in Figure \ref{fig:bce}. It is immediately clear that the two unweighted models perform much worse than the weighted models. These unweighted models do not account for the informative sample design and thus introduce a large amount of bias when making inference on the population. The weighted models are able to account for the sample design, and thus result in much lower values of BCE. Additionally, the distribution of BCE under FM-W is shifted to the left of SM-W, indicating that the functional covariate does aid in prediction of mortality for member of the population. These results indicate that for population level inference based on the full sample data, we should use the model that utilizes functional covariates while also accounting for the survey design through a Bayesian pseduo-likelihood.

\section{NHANES Data Analysis}\label{sec: DA}

\subsection{5-Year Mortality Estimate}

Using our Bayesian pseudo-likelihood based model for functional covariates (FM-W), we now analyze the NHANES PAM data and its relationship with mortality. We use the same dataset considered in the simulation study and outlined in Section \ref{sec: data}, with a sample size of 3,208. We treat the 5-year mortality indicator as our binary response, and use an intercept and age as our scalar covariates as well as the PAM data as a functional covariate. We use the FACE functional principal components basis representation and retain the first 19 components, explaining 95\% of the variation in the functional data.

We fit the model via Gibbs sampling with 5,000 iterations and discard the first 1,000 iterations as burn-in. Convergence was assessed via traceplots of the sample chains, where no lack of convergence was detected.

\begin{figure}[H]
    \begin{center}
        \includegraphics[width=150mm]{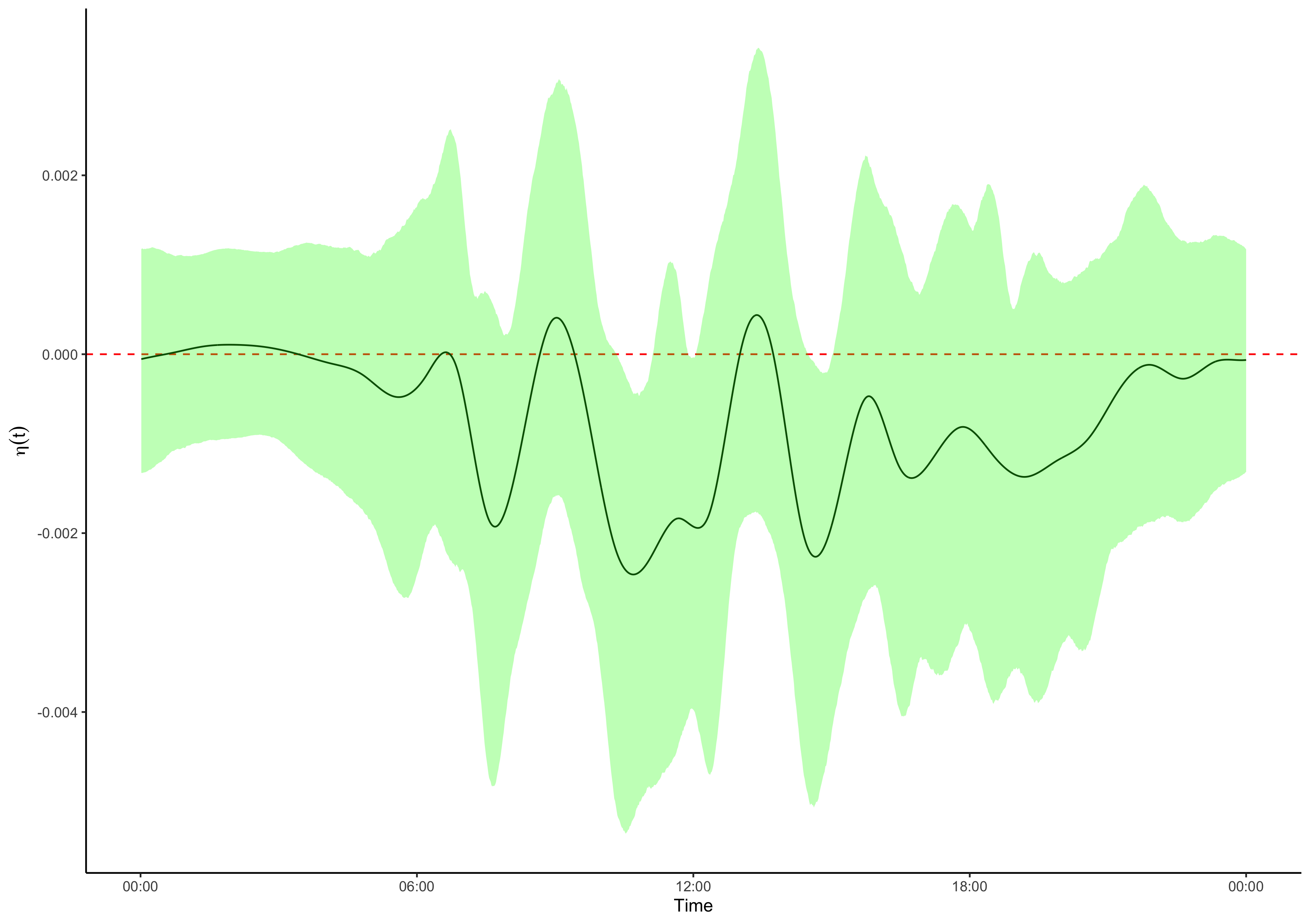}
        \caption{\baselineskip=10pt Estimate of the PAM functional regression coefficient for 5-year mortality along with pointwise 90\% credible interval.}
        \label{fig: reg}
    \end{center}               
\end{figure}

After fitting the model, we are able to make population level inference. We plot the posterior mean of the functional regression coefficient, $\eta(t)$, along with a pointwise 90\% credible interval in Figure \ref{fig: reg}. For the most part the $\eta(t)$ is estimated to be negative, as expected, indicating that increased levels of activity are associated with lower expected mortality rate. There are two primary time periods where the credible interval does not contain zero, around 10 a.m. to 12 p.m. and around 2 p.m. to 3 p.m.

\begin{figure}[H]
    \begin{center}
        \includegraphics[width=150mm]{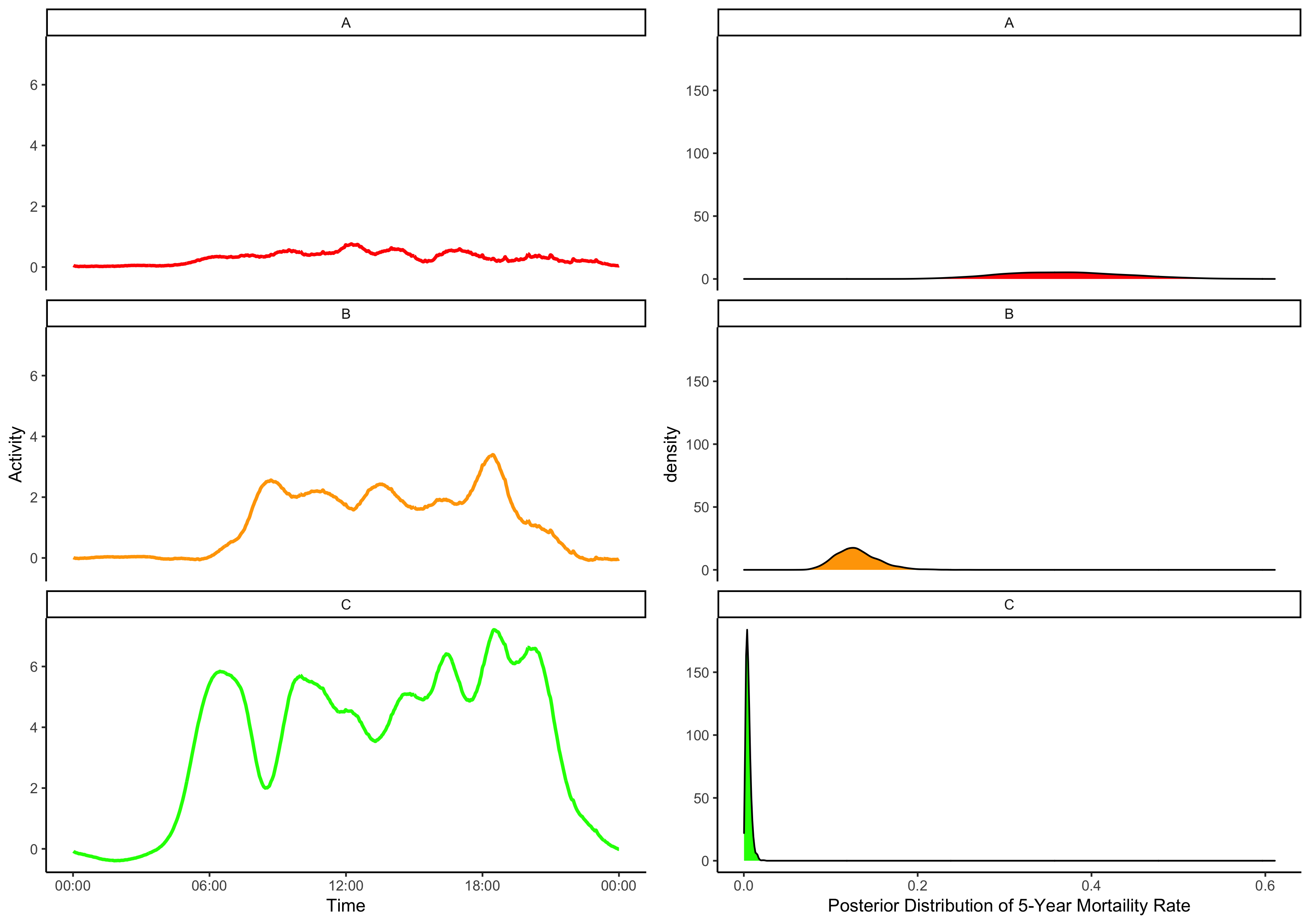}
        \caption{\baselineskip=10pt  Activity curves for 3 hypothetical individuals all age 65 along with accompanying posterior distribution of 5-year mortality rate.}
        \label{fig: 5yr}
    \end{center}               
\end{figure}

In addition to examination of the functional regression coefficient, we can also glean insight by examining how variation in activity level of individuals changes mortality estimates. In Figure \ref{fig: 5yr}, we plot activity curves for 3 hypothetical individuals all age 65 along with the accompanying posterior distribution of 5-year mortality rate. Individual A has a very low level of activity, resulting in a high expected mortality rate. However, there is also a great deal of uncertainty around this rate. Individuals B and C both have increasing level of overall activity, especially in the early morning and late afternoon, resulting in decreasing expected mortality rate. As the activity level increases, the uncertainty around the mortality rate decreases.

\subsection{Multivariate Mortality Estimate}
In addition to univariate estimation, our model allows for multivariate estimation through the Multinomial data model. In this case we wish to make joint mortality estimates for years 1-5. To do so, we begin by assigning survey respondents into distinct categories: those who died within one year of the survey, those who died after 1 year but before 2 years, those who died after 2 years but before 3 years, those who died after 3 years but before 4 years, those who died after 4 years but before 5 years, and finally those that did not die before 5 years. Assigning groups in this way results in a Multinomial or Categorical data distribution with 6 categories. Thus, we are able to use the stick breaking representation of the Multinomial distribution in order to fit $C-1=5$ independent Binomial data models that allow us to make joint estimates of mortality at various time points.

\begin{figure}[H]
    \begin{center}
        \includegraphics[width=150mm]{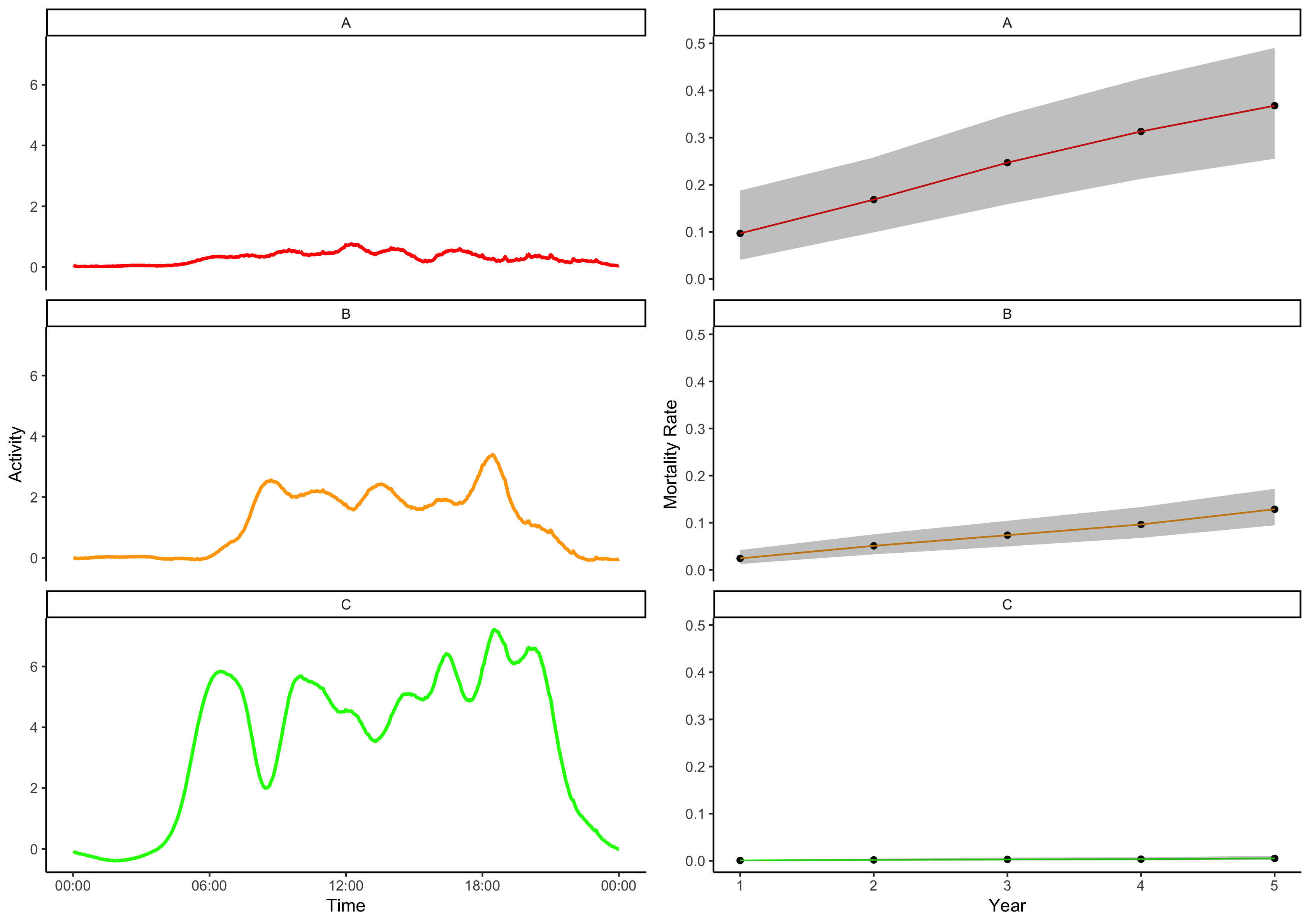}
        \caption{\baselineskip=10pt  Activity curves for 3 hypothetical individuals all age 65 along with accompanying posterior mean mortality rate for years 1-5 and 90\% credible intervals.}
        \label{fig: joint}
    \end{center}               
\end{figure}

We use the same hypothetical individuals from our 5-year mortality example to examine the effects of activity level on multi-year mortality. Figure \ref{fig: joint} plots the activity curves for these individuals alongside their posterior mean mortality rates for year 1-5. We also provide 90\% credible intervals. Once again, we see that both expected mortality rate and uncertainty increase as activity level generally decreases. Because these estimates are multivariate, we can also see that decreased activity is associated with steeper marginal increases in mortality for the near future than for years further away from the survey.

\section{Discussion}\label{sec: disc}
In this work, we develop a Bayesian non-Gaussian data model for functional covariates under informative sampling. We rely on a pseudo-likelihood approach to account for survey design which works in combination with Pólya-Gamma data augmentation to allow for conjugate full conditional distributions of the regression parameters. This method is designed for Binomial or Multinomial data models, though it is straightforward to replace this with a Gaussian data model. Our approach uses an orthonormal basis representation of the functional covariates alongside the Horseshoe prior to provide regularization. As with the data model, we use a data augmentation approach for the Horsehoe prior, meaning that all full-conditional distributions in the model are conjugate. This allows for straightforward and efficient Gibbs sampling, which can be highly important in high-dimensional settings such as the one explored here.

We conduct an empirical simulation study using NHANES data that shows that our approach is able to reduce the bias attributable to informative sampling while also making use of the functional data to improve estimates for members of the population. We also provide a full analysis of the NHANES data that allows us to make inference and prediction on the population. We conduct both a univariate analysis concerning 5-year mortality rate as well as a multivariate analysis concerning years 1-5 mortality rate.

Our methodology extends the literature on functional regression to the survey data setting. The approach is flexible in that users have a choice of data model and basis expansion and also allows for joint estimation of scalar regression coefficients. Currently, there is a limited amount of functional data collected under complex surveys, as analysis options are limited. It is our hope that with the availability of this methodology, collection of functional data via surveys will become more widespread.

Although not explored in this work, similar approaches may be undertaken for function on scalar or function on function regression under complex survey designs. Another potential avenue of future research would involve the use of nonlinear modeling techniques that utilize these same functional covariates. Finally, although we were able to utilize multivariate techniques to jointly estimate mortality at multiple time points, it would be interesting to explore models that can estimate continuous survival curves based on the NHANES activity data.
\section*{Acknowledgement}
Support
  for this research through the Census Bureau Dissertation Fellowship program is gratefully acknowledged. This research was partially supported by the
  U.S.~National Science Foundation (NSF) under NSF grant SES-1853096. This article is released to inform interested parties of ongoing
  research and to encourage discussion. The views expressed on statistical issues are
  those of the authors and not those of the NSF or U.S. Census Bureau.

\clearpage\pagebreak\newpage

\baselineskip=14pt %\vskip 2mm\noindent
\bibliographystyle{jasa}
\bibliography{nhanes}

\section*{Appendix: Full Conditional Distributions}

Let $\bm{\Omega} = \hbox{diag}(\omega_1,\ldots,\omega_n)$, $\bm{\Lambda}=\hbox{diag}(\lambda_1^2, \ldots, \lambda_K^2)$, and $\bm{\kappa}=\left(\tilde{w}_1*(y_1 - n_1/2), \ldots, \tilde{w}_n*(y_n - n_n/2) \right)'$. Note that $\bm{\kappa}/\bm{\omega}$ represents element-wise division.

%\begin{equation}
    $$\begin{aligned}
    \omega_i | \cdot & \sim \hbox{PG}(\tilde{w}_i*n_i, \; \bm{x}_i'\bm{\beta} + \sum_{k=1}^{K}b(k)\xi_i(k)), \; i=1,\ldots,n \\
            \bm{b} | \cdot & \propto  \prod_{i=1}^n \hbox{exp}\left(\kappa_i \bm{\xi}_i'\bm{b} - \frac{1}{2}\omega_i(\bm{\xi}_i'\bm{b})^2 - \omega_i (\bm{\xi}_i'\bm{b})(\bm{x}_i'\bm{\beta})\right) \\
            & \times \hbox{exp}\left(-\frac{1}{2\tau^2}\bm{b}'\bm{\Lambda}^{-1}\bm{b}\right) \\
            & \propto \hbox{exp}\left(-\frac{1}{2}(\bm{\kappa}/\bm{\omega} - \bm{X\beta} - \bm{\Xi}\bm{b} )' \bm{\Omega} (\bm{\kappa}/\bm{\omega}  - \bm{X\beta} - \bm{\Xi}\bm{b})-\frac{1}{2\tau^2}\bm{b}'\bm{\Lambda}^{-1}\bm{b} \right) \\
        \bm{b} | \cdot & \sim \hbox{N}_K\left(\bm{\mu} = (\bm{\Xi'\Omega \Xi} + \frac{1}{\tau^2} \bm{\Lambda}^{-1} )^{-1}\bm{\Xi}'\bm{\Omega}(\bm{\kappa}/\bm{\omega}-\bm{X \beta}), \;  \bm{\Sigma}=(\bm{\Xi}'\bm{\Omega \Xi} + \frac{1}{\tau^2} \bm{\Lambda}^{-1})^{-1}   \right) \\
    \end{aligned}$$

     $$\begin{aligned}
        \bm{\beta} | \cdot & \propto  \prod_{i=1}^n \hbox{exp}\left(\kappa_i \bm{x}_i'\bm{\beta} - \frac{1}{2}\omega_i(\bm{x}_i'\bm{\beta})^2 - \omega_i (\bm{x}_i'\bm{\beta})(\bm{\xi}_i'\bm{b})\right) \\
        & \times \hbox{exp}\left(-\frac{1}{2\sigma^2_{\beta}}\bm{\beta}'\bm{\beta}\right) \\
        & \propto \hbox{exp}\left(-\frac{1}{2}(\bm{\kappa}/\bm{\omega} - \bm{\Xi}\bm{b} - \bm{X\beta})' \bm{\Omega} (\bm{\kappa}/\bm{\omega} - \bm{\Xi}\bm{b} - \bm{X\beta}) - \frac{1}{2\sigma^2_{\beta}}\bm{\beta}'\bm{\beta} \right) \\
        \bm{\beta} | \cdot & \sim \hbox{N}_q\left(\bm{\mu} = (\bm{X'\Omega X} + \frac{1}{\sigma^2_{\beta}} \bm{I}_q)^{-1}\bm{X'}\bm{\Omega}(\bm{\kappa}/\bm{\omega}-\bm{\Xi b}), \;  \bm{\Sigma}=(\bm{X'\Omega X} + \frac{1}{\sigma^2_{\beta}} \bm{I}_q)^{-1}   \right) \\
    \end{aligned}$$
    
    $$\begin{aligned}
        \lambda_k^2 | \cdot & \propto (\lambda_k^2)^{-1/2}\hbox{exp}\left(-\frac{b(k)^2}{2\tau^2\lambda_k^2} \right) \\
        & \times (\lambda_k^2)^{-3/2} \hbox{exp}\left(-\frac{1}{\nu_k\lambda_k^2} \right) \\
        & \propto (\lambda_k^2)^{-2} \hbox{exp}\left\{-\frac{1}{\lambda_k^2}\left(\frac{1}{\nu_k} + \frac{b(k)^2}{2\tau^2} \right) \right\} \\
        \lambda_k^2 | \cdot & \sim \hbox{IG}\left(1, \frac{1}{\nu_k} + \frac{b(k)^2}{2\tau^2} \right)
    \end{aligned}$$
    
        $$\begin{aligned}
        \tau^2 | \cdot & \propto (\tau^2)^{-K/2}\hbox{exp}\left(-\frac{1}{\tau^2}\sum_{k=1}^K\frac{b(k)^2}{2\lambda_k^2} \right) \\
        & \times (\tau^2)^{-3/2} \hbox{exp}\left(-\frac{1}{\nu_{\tau}\tau^2} \right) \\
        & \propto (\tau^2)^{-\frac{K+1}{2}-1} \hbox{exp}\left\{-\frac{1}{\tau^2}\left(\frac{1}{\nu_{\tau}} + \sum_{k=1}^K\frac{b(k)^2}{2\lambda_k^2} \right) \right\} \\
        \tau^2 | \cdot & \sim \hbox{IG}\left(\frac{K+1}{2}, \frac{1}{\nu_{\tau}} + \sum_{k=1}^K\frac{b(k)^2}{2\lambda_k^2} \right)
    \end{aligned}$$
    
    $$\begin{aligned}
    \nu_k | \cdot & \propto \nu_k^{-3/2} \hbox{exp}\left(-\frac{1}{\nu_k} \right) \\ 
    & \times \nu_k^{-1/2} \hbox{exp}\left(-\frac{1}{\nu_k\lambda_k^2}\right) \\
    & \propto \nu_k^{-2} \hbox{exp}\left\{-\frac{1}{\nu_k} \left(1 + \frac{1}{\lambda_k^2} \right) \right\} \\
    \nu_k | \cdot & \sim \hbox{IG}\left(1, 1 + \frac{1}{\lambda_k^2} \right)
    \end{aligned}$$
    
        $$\begin{aligned}
    \nu_{\tau} | \cdot & \propto \nu_{\tau}^{-3/2} \hbox{exp}\left(-\frac{1}{\nu_{\tau}} \right) \\ 
    & \times \nu_{\tau}^{-1/2} \hbox{exp}\left(-\frac{1}{\nu_{\tau}\tau^2}\right) \\
    & \propto \nu_{\tau}^{-2} \hbox{exp}\left\{-\frac{1}{\nu_{\tau}} \left(1 + \frac{1}{\tau^2} \right) \right\} \\
    \nu_{\tau} | \cdot & \sim \hbox{IG}\left(1, 1 + \frac{1}{\tau^2} \right)
    \end{aligned}$$

%\end{equation}

\end{document}